\documentclass[reprint,superscriptaddress,amssymb,amsmath,sn,prxq]{revtex4-2}

\usepackage{floatrow}
\usepackage{graphicx}
\usepackage{amssymb}
\usepackage{epstopdf}
\usepackage{upgreek}
\usepackage{dcolumn}
\usepackage{bm}
\usepackage{gensymb}
\usepackage{braket}
\usepackage{amsmath}
\usepackage{mathtools}
\usepackage{float}
\usepackage{tikz}
\usepackage[colorlinks=true,allcolors=blue]{hyperref}

\usepackage{array}
\newcolumntype{C}[1]{>{\centering\arraybackslash}m{#1}}

\usepackage{xcolor}

\usepackage[resetlabels,labeled]{multibib}

\usepackage{xfrac}

\usepackage{comment}

\usepackage{hhline}

\usepackage[showframe=false]{geometry}
\usepackage{enumitem}

\usepackage{physics}

\usepackage{lmodern}
\usepackage{microtype}

\allowdisplaybreaks

\DeclareMathOperator{\arccosh}{arccosh}

\expandafter\ifx\csname package@font\endcsname\relax\else
\expandafter\expandafter
\expandafter\usepackage
\expandafter\expandafter
\expandafter{\csname package@font\endcsname}%
\fi
\newcommand{\ie}{\emph{i.e.}}

\newcommand{\micro}{$\upmu$}
\newcommand{\uA}{$\upmu$A}

\newcommand{\be}{\begin{eqnarray}}
\newcommand{\ee}{\end{eqnarray}}
\newcommand{\bfig}{\begin{figure}}
	\newcommand{\efig}{\end{figure}}

\definecolor{myorange}{rgb}{0.97,0.59,0.27}

\newcommand{\cdotscompact}{\cdots\!}

\newcommand{\cor}[1]{#1}

\DeclareFontFamily{U}{mathb}{}
\DeclareFontShape{U}{mathb}{m}{n}{
	<-5.5> mathb5
	<5.5-6.5> mathb6
	<6.5-7.5> mathb7
	<7.5-8.5> mathb8
	<8.5-9.5> mathb9
	<9.5-11.5> mathb10
	<11.5-> mathbb12
}{}

\begin{document}
\makeatletter
  \@namedef{figure}{\killfloatstyle\def\@captype{figure}\FR@redefs
    \flrow@setlist{{figure}}%
    \columnwidth\columnwidth\edef\FBB@wd{\the\columnwidth}%
    \FRifFBOX\@@setframe\relax\@@FStrue\@float{figure}}%
\makeatother

\setlist[itemize]{wide = 0pt}

\title{Artificial Transmission Line Synthesis Tailored for Traveling-Wave Parametric Processes}

\author{M. Malnou}
\email{maxime.malnou@nist.gov}
\affiliation{National Institute of Standards and Technology, 325 Broadway, Boulder, CO 80305, USA}
\affiliation{University of Colorado, 2000 Colorado Ave., Boulder, CO 80309, USA}

\date{\today}

\begin{abstract} 

Artificial transmission lines built with lumped-element inductors and capacitors form the backbone of broadband, nearly quantum-limited traveling-wave parametric amplifiers (TWPAs). \cor{When tailoring these transmission lines for parametric processes, nonlinear elements are added, typically nonlinear inductances in superconducting circuits, and energy and momentum conservation between interacting tones must be enforced through careful design of the ATL dispersion relation. However, a unified theoretical framework describing achievable dispersion relations is lacking. Here, I develop such a framework, borrowing from periodic structure theory and passive network synthesis. These complementary approaches divide the design space: periodic loading synthesis employs spatial modulation of frequency-independent components, while filter synthesis employs frequency-dependent responses in spatially-uniform components. The framework reveals fundamental constraints and enables the discovery of novel TWPA architectures. In particular, I design a kinetic inductance TWPA with a novel phase-matching architecture, and a backward-pumped Josephson TWPA exploiting an ambidextrous \ie{}, right-left-handed transmission line.}

\end{abstract}

\maketitle

\section{Introduction}

In linear circuit design, synthesis and analysis are complementary. Analysis computes the response of a given network, for example by cascading ABCD matrices \cite{pozar2011microwave}, whereas synthesis tackles the inverse problem: designing a network that approximates a desired response \cite{guillemin1957synthesis}. Though synthesis does not guarantee the resulting network is unique or optimal, \cor{it is built on a theoretical framework that reveals what responses are achievable and what network topologies are possible, enabling the exploration of exotic device concepts.}

While resonant parametric amplifiers have recently benefited from systematic synthesis approaches for gain and bandwidth \cite{naaman2022synthesis}, traveling-wave parametric amplifiers (TWPAs), essential components in superconducting quantum circuit readout \cite{andersen2020Repeated,krinner2022realizing,malnou2023improved,vora2023Search,bartram2023dark}, lack equivalent design tools. This absence stems partly \cor{from the lack of such a framework describing the dispersion of} artificial transmission lines (ATLs) that form the backbone of TWPA architectures.

\cor{Controlling the ATL's dispersion relation is essential} when designing a TWPA, because a key challenge is to selectively enable mode propagation in order to favor desired parametric processes, while suppressing spurious processes that deplete pump photons and degrade noise performance \cite{remm2023intermodulation,pengPRXQuantumFloquet}. A phase-mismatch will prevent the exponential growth of a given process along the ATL, \cor{but to effectively suppress it, a stopband can be engineered to prevent the coherent buildup of one of the modes involved}. Mode filtering becomes all the more critical in devices leveraging several parametric processes, for example parametric amplification and frequency conversion \cite{malnou2025a,ranadive2024a}, where multiple pump tones can mix together \cor{and produce larger sets of spurious processes}. \cor{Heuristic} mitigation strategies exist, ranging from leveraging natural chromatic dispersion \cite{macklin2015near,white2015traveling,qiu2023broadband}, to periodic loading techniques \cite{Eom2012a,vissers2016low,malnou2021three,gaydamachenko2022numerical,gaydamachenko2025rf,howe2025kinetic} and Floquet mode propagation \cite{pengPRXQuantumFloquet,wang2025high}.

Here, I develop a \cor{theoretical framework} for lossless ATLs by recognizing that the design space is divided into two complementary \cor{synthesis} approaches: periodic loading, where frequency-independent components vary spatially, and filter synthesis, where spatially-invariant components exhibit frequency-dependent responses. In Sec.\,\ref{sec:ATLsynth}, I present this theoretical framework, developing both approaches and their fundamental constraints. In Sec.\,\ref{sec:examples}, I apply these methods to design two TWPAs: a kinetic inductance device with a novel phase-matching structure, and a backward-pumped Josephson TWPA exploiting ``ambidextrous'' transmission line behavior. I conclude by discussing the broader implications of \cor{this framework} for superconducting quantum circuits.

\section{Artificial Transmission Line Synthesis}
\label{sec:ATLsynth}

\subsection{General form for an artificial transmission line}
\begin{figure}[h!]
\includegraphics[width=\columnwidth]{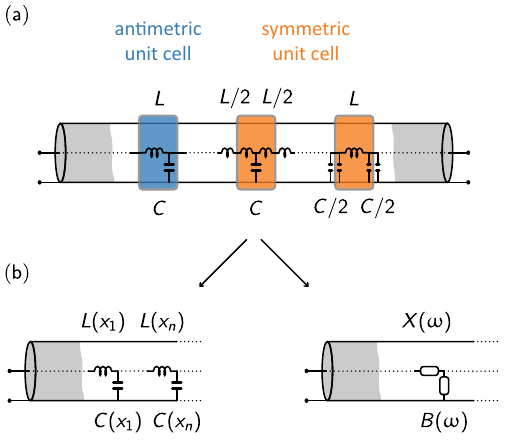}
\caption{General topology of \cor{a right-handed} artificial transmission line. (a) Cascaded unit cells, comprising series reactances (typically inductors) and shunt susceptances (typically capacitors), can be made antimetric (blue) or symmetric (orange). (b) The synthesis framework divides into two approaches: periodic loading (left) where the inductors and capacitors vary spatially, and filter synthesis (right) where the spatially-invariant reactances (X) and susceptances (B) are engineered to give a specific frequency response.}
\label{fig:ATL}
\end{figure}

In the context of electrical circuits, an ATL is a one-dimensional periodic lattice built with lumped circuit elements, supporting the propagation of voltage and current waves. Topologically, a lossless, \cor{right-handed} ATL consists of an infinite cascade of series reactances, typically inductors, alternating with susceptances, typically capacitors, shunted to ground. A unit cell within the ATL is composed of an inductor-capacitor pair; it is antimetric when the entire inductor precedes or succeeds the capacitor, and symmetric when either the inductor or the capacitor is halved and surrounds the other element within the pair, see Fig.\,\ref{fig:ATL}a. Both inductors and capacitors can have a periodic spatial dependency. Both elements can also be replaced with sub-networks of inductors and capacitors. This leads to two ways of designing the response of an ATL, see Fig.\,\ref{fig:ATL}b: engineer the periodic modulation of inductors and capacitors, or engineer the frequency response of spatially-invariant reactances and susceptances. Below, I present a synthesis framework for both approaches separately, and then discuss the effect of combining them within the same lattice.

\subsection{Periodic loading synthesis}

Spatially modulating frequency-independent components creates stopbands in the ATL response whose systematic design requires a generalized dispersion relation that captures the effect of periodic loading. I derive this relation in Sec.\,\ref{sec:disprelpl}, then show in Sec.\,\ref{sec:stopbandsynth} how to synthesize stopbands by specifying their center frequencies and widths. \cor{Here, the right-handed ATL can be viewed as a prototype; other topologies, including left-handed and composite lines, can be obtained by applying the frequency transformations of Sec.\,\ref{sec:freq_trans} to this prototype (see Sec.\,\ref{sec:topo_considerations}).}

\subsubsection{Dispersion relation}
\label{sec:disprelpl}

The response of a lossless, infinite ATL is described by its dispersion relation $k(\omega)$, where $k$ is the ATL's wavenumber and $\omega$ the frequency. When $L$ and $C$ are constant, $k^2=\omega^2LC$. The goal here is to find the dispersion relation of an ATL whose series inductor $L(x)$ and shunt capacitor $C(x)$ can vary periodically as a function of the position $x$ along the line. Assuming that the ATL is periodic over $d$ unit cells \ie{}, $d$ defines the length of a supercell \cite{malnou2021three}, $L$ and $C$ can be expanded in spatial Fourier harmonics:
\begin{align}
    L(x) &= L_0\sum_{n=-\infty}^{\infty} l_n e^{-j n k_d x} \label{eq:LFourier}\\
    C(x) &= C_0\sum_{n=-\infty}^{\infty} c_n e^{-j n k_d x}, \label{eq:CFourier}
\end{align}
where $k_d = 2\pi/d$. Note that since $L$ and $C$ are real\cor{-valued functions}, \cor{$l_n=l_{-n}^*$} and \cor{$c_n=c_{-n}^*$} for $n\in\mathbb{Z}$.

The voltage $V(x,t)$ and current $I(x,t)$ at position $x$ and time $t$ can be decomposed onto the same basis. In fact, for a sinusoidal signal $V(x,t)=V(x)e^{j\omega t}$ and $I(x,t)=I(x)e^{j\omega t}$ with $j^2=-1$, Bloch's theorem \cite{Bloch1929uber} states that both $V(x)$ and $I(x)$ are periodic functions with the same lattice periodicity $d$, modulated by a phase $e^{-\gamma x}$. Here, $\gamma=\alpha+j k$ is the propagation constant, and in the passband region of the lossless ATL, $\alpha=0$ therefore $\gamma =j k$, which yields
\begin{align}
    V(x) &= e^{-jkx}\sum_{n=-\infty}^{\infty} v_n e^{-j n k_d x} \label{eq:VFourier} \\
    I(x) &= e^{-jkx}\sum_{n=-\infty}^{\infty} i_n e^{-j n k_d x}. \label{eq:IFourier}
\end{align}
Note that $V$ and $I$ contain both forward and backward propagating waves, corresponding to $k + nk_d > 0$ and $k + nk_d < 0$, respectively.

The dispersion relation follows from using Eqs.\,\ref{eq:LFourier}-\ref{eq:IFourier} in the Telegrapher's propagation equations. In the frequency domain, and for an antimetric unit cell, these equations govern the evolution of $V$ and $I$:
\begin{align}
    \frac{dV}{dx} &= -j\omega L(x) I(x) \label{eq:dV}\\
    \frac{dI}{dx} &= -j\omega C(x) V(x) \label{eq:dI}.
\end{align}
Replacing Eqs.\,\ref{eq:LFourier}-\ref{eq:IFourier} into Eqs.\,\ref{eq:dV} and \ref{eq:dI} then yields:
\begin{align}
    (k + nk_d)v_n &+ \omega L_0 \sum_m l_{n-m}\cor{i_m} = 0 \\
    (k + nk_d)i_n &+ \omega C_0 \sum_m c_{n-m}\cor{v_m} = 0,
\end{align}
for $n\in\mathbb{Z}$. Normalizing the impedance to $Z_b=\sqrt{L_0/C_0}=1$\,\ohm{}, the frequency to $\omega_0=1/\sqrt{L_0C_0}=1$\,rad/s, and defining $\Tilde{k}=k/k_d$ and $\Tilde{\omega}=\omega/k_d$ leads to the system of equations
\begin{align}
    &\mathbf{K} \mathbf{v} + \tilde{\omega} \mathbf{L}\mathbf{i} = \mathbf{0} \label{eq:vecdv}\\
    &\mathbf{K} \mathbf{i} + \tilde{\omega} \mathbf{C}\mathbf{v} = \mathbf{0}, \label{eq:vecdi} 
\end{align}
where $\mathbf{K}=\text{diag}(\tilde{k}+n)$, $\mathbf{L} = [l_{n-m}]_{n,m}$, $\mathbf{C} = [c_{n-m}]_{n,m}$, $\mathbf{v} = [v_n]^T$, and $\mathbf{i} = [i_n]^T$ for $\{n,m\}\in\mathbb{Z}$. This system has a non-trivial solution when
\begin{equation}
    D\triangleq\det(\tilde{\omega}^2\mathbf{L}\mathbf{C}-\mathbf{K}^2)=0, \label{eq:disprel}
\end{equation}
which is the generalized dispersion relation of a periodically loaded, \cor{right-handed} ATL. Here, $\mathbf{M}=\tilde{\omega}^2\mathbf{L}\mathbf{C}-\mathbf{K}^2$ defines the coupling matrix between the spatial harmonics $\{e^{-j n k_d x}\}_{n\in\mathbb{Z}}$.

\subsubsection{Stopband synthesis}
\label{sec:stopbandsynth}

\begin{figure*}[hptb!]
\includegraphics[width=\textwidth]{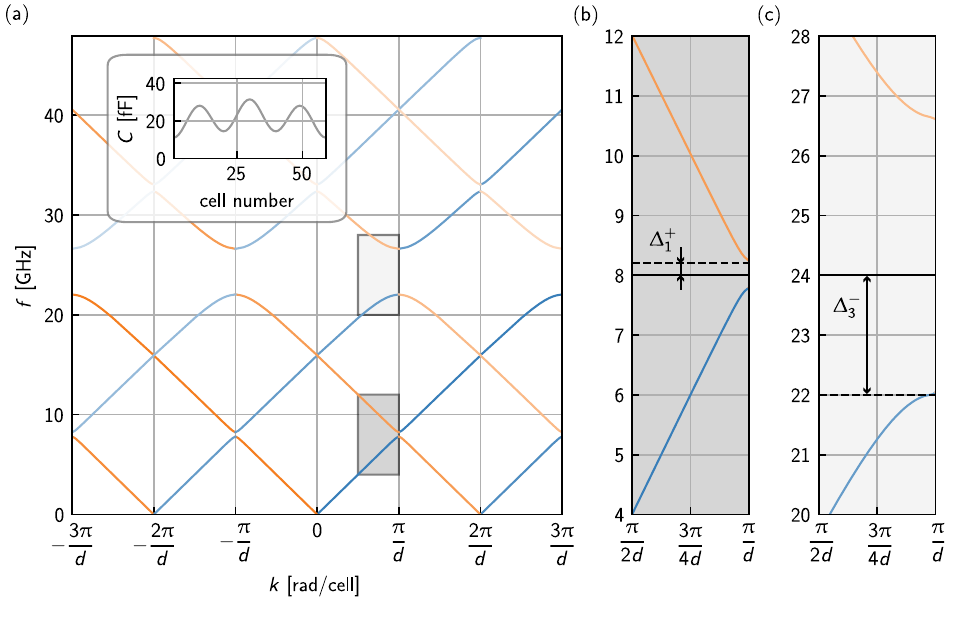}
\caption{Example of stopband synthesis via periodic modulation of the ATL shunt capacitance. (a) The band diagram shows the dispersion relation for the periodically loaded ATL with two synthesized stopbands at $8$ and $24$\,GHz. Each line represents a different spatial harmonic $n$, with blue lines indicating modes where $k + nk_d > 0$ (forward propagation) and orange lines where $k + nk_d < 0$ (backward propagation). The stopbands appear at the edge of the Brillouin zones where counter-propagating modes couple, creating the bandgaps shown in gray. The inset shows the capacitance modulation profile over one supercell ($d = 59$ cells), consisting of two overlapped cosine modulations. (b) Zoom of the first stopband at $8$\,GHz with width $\Delta_1^+ = 0.2$\,GHz. (c) Zoom of the third stopband at $24$\,GHz with width $\Delta_3^- = 2$\,GHz.}
\label{fig:plsynth}
\end{figure*}

Modulating $L$ and $C$ creates stopbands in the ATL's transmission as a function of frequency, whose position and width can be synthesized using the general dispersion relation. But if the modulations of $L$ and $C$ are identical, the ATL is equivalent to an unloaded transmission line: there is no stopband, see appendix \ref{app:nomodATL}. In other words, only the periodic modulation of the ATL's Bloch impedance $Z_b=\sqrt{L/C}$ creates stopbands. Thus, to simplify the discussion, we modulate $Z_b$ by fixing $L=L_0=1$ (in normalized units) and place all spatial dependence in $C$. In that case, the dispersion relation (Eq.\,\ref{eq:disprel}) reads \cite{hwang2012periodic}
\begin{equation}
\begin{aligned}
&D =\\
&\setlength{\arraycolsep}{2.5pt}
\begin{vmatrix*}
\vdots & \vdots & \ddots & \vdots & \vdots \\
\cdotscompact & \tilde{\omega}^2-(\tilde{k}-1)^2 & \tilde{\omega}^2 c_1 & \tilde{\omega}^2 c_2 & \cdotscompact \\
\cdotscompact & \tilde{\omega}^2 \cor{c_1^*} & \tilde{\omega}^2-\tilde{k}^2 & \tilde{\omega}^2 c_1 & \cdotscompact \\
\cdotscompact & \tilde{\omega}^2 \cor{c_2^*} & \tilde{\omega}^2 \cor{c_1^*} & \tilde{\omega}^2-(\tilde{k}+1)^2 & \cdotscompact \\
\vdots & \vdots & \ddots & \vdots & \vdots
\end{vmatrix*}\\
&= 0.
\end{aligned}
\label{eq:det}
\end{equation}

The problem of synthesis then consists of imposing that the dispersion relation pass through $N$ points defined by the pairs $\{\tilde{k}_n,\tilde{\omega}_n\}_{n\in N}$, and to find the Fourier amplitudes $\{c_n\}_{n\in N}$ that fulfill such constraints. Each constraint generates a different determinant $D_n$, and their ensemble then forms a system of polynomial equations $\{D_n=0\}_{n\in N}$ whose variables are the Fourier amplitudes $\{c_n\}_{n\in N}$.

In a low-pass ATL where the unit cell is composed of a series inductor and a shunt capacitor, stopbands can only exist at the edge of the Brillouin zones (and not at zero frequency), see Sec.\,\ref{sec:kfoster}. In normalized units, this corresponds to $\tilde{k}=n/2$, for $n\in\mathbb{Z} \setminus \{0\}$. Furthermore, without periodic modulation $\tilde{k}=\tilde{\omega}$. So, in the context of stopband synthesis, the constraints take the form  $\{\tilde{k}_n=n/2,\tilde{\omega}_n=n/2\pm\delta_n^{\pm}\}$, where $\pm\delta_n^{\pm}$ is the upper or lower stopband width.

Even though $\mathbf{M}$ is an infinite matrix, it can be truncated because the stopband structure is fully determined by the coupling between harmonics in the range $n \in [-N, N]$, where $N$ corresponds to the highest stopband frequency. When designing several stopbands, $\tilde{k}_1=1/2$ is associated to the greatest common divisor (gcd) frequency $\tilde{\omega}_1$ of all the stopband frequencies. In other words, \cor{the number $d$ of cells per supercell} is chosen so that all the stopband frequencies can be decomposed onto the same spatial harmonic basis. In that case, $\mathbf{M}$ may be sparse, because only the harmonics corresponding to a stopband create non-zero off-diagonal terms.

As an example, Fig.\,\ref{fig:plsynth} shows the synthesized band diagram of an ATL, suitable for 4WM amplification \cite{Eom2012a}: the pump frequency is placed slightly below a first stopband at $8$\,GHz, while another, wide stopband at $24$\,GHz cuts the third-harmonic generation process \cite{Landauer1960shock,boyd2019nonlinear}. Here, the gcd $f_1=8$\,GHz and $\dim(\mathbf{M})=7$, but the second super- and sub-diagonals of $\mathbf{M}$ only contain zeros. The stopband widths are asymmetric; the upper ($+$, above center frequency) and lower ($-$, below) widths are set to $\Delta_1^+=0.2$\,GHz and $\Delta_3^-=2$\,GHz. The inset in Fig.\,\ref{fig:plsynth}a shows the normalized capacitance modulation profile over one supercell. It consists of two overlapped cosine modulations.

The constraints generating $\{D_n=0\}_{n\in N}$ may not always yield realizable solutions. In fact, physical realizability requires $\lvert c_n \rvert \leq 0.5$ (necessary but not generally sufficient), because $C$ cannot take negative values (since \cor{$C=C_0[1+c_1e^{-jk_dx} + c_1^*e^{jk_dx} + \cdots$]}, see Eq.\,\ref{eq:CFourier}). In other words, the stopbands have a maximum width. For the case of a single stopband at $\tilde{k}=1/2$, where only the $n=0$ and $n=-1$ harmonics couple, the relative stopband width $\delta_1=(\Delta_1^++\Delta_1^-)/f_1$ can be derived analytically; $\delta_1^{\mathrm{max}} \simeq 0.6$ when $c_1=0.5$, see appendix \ref{app:stopwidth}.

Looking at Fig.\,\ref{fig:plsynth}a, the k-th stopband mainly originates from the coupling of two counter-propagating modes, one at $n=0$ and one at $n=-k$. Nonetheless, each stopband cannot be synthesized independently. In other words, the coefficients $c_1$ and $c_3$ found by solving the system of two determinants $\{D_1=0,D_3=0\}$ are different from that found by solving $D_1=0$ separately for the two situations, \ie{} when $\tilde{\omega}_1=1/2$ corresponds to $8$\,GHz or $24$\,GHz. This interdependence comes from the fact that the determinant equations are nonlinear, in other words each stopband influences the dispersion relation, even far from the stopband itself.

\subsection{Filter synthesis for ATL}
\label{sec:filtsynth}

I now adapt filter synthesis techniques to design the response of an ATL. Any reactance and susceptance is allowed, but all the unit cells within the ATL are identical. In filter synthesis, starting from a low-pass filter prototype, there exist systematic transformations to convert this prototype into filters comprising any number of passbands and stopbands \cite{guillemin1957synthesis}. I apply this framework to ATL synthesis by establishing a parallel: in Sec.\,\ref{sec:LPATL}, I construct a low-pass ATL prototype, and in Sec.\,\ref{sec:freq_trans} I transform this prototype to create ATLs with arbitrary pass-stop responses.

\subsubsection{Low-pass ATL prototype}
\label{sec:LPATL}

In a low-pass filter (LPF) prototype, the components are arranged in an LC ladder network of series inductors and shunt capacitors, see appendix \ref{app:filter_synth}. The low-pass ATL, sometimes called \textit{right-handed} \cite{caloz2005electromagnetic}, follows the same structure, with a LPF as its unit cell, see Fig.\,\ref{fig:ATL}a.

\begin{figure}[h!]
\includegraphics[width=\columnwidth]{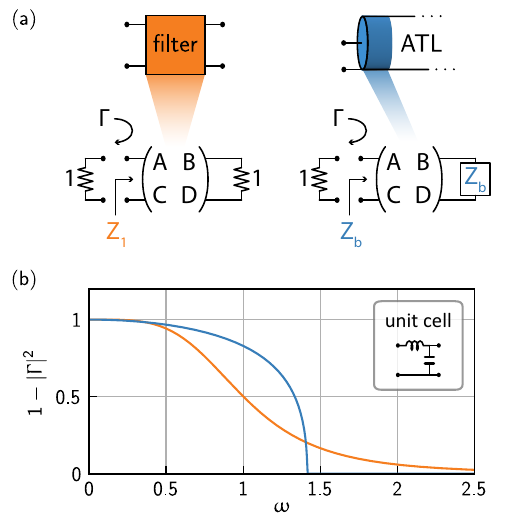}
\caption{Concept of a low-pass ATL prototype. (a) Both filters and ATLs can be characterized by their reflection coefficient $\Gamma$. For a filter, $\Gamma$ relates to the input impedance $Z_1$ computed from the ABCD matrix with a $1$\,\ohm{} termination. For an ATL, $\Gamma$ relates to the Bloch impedance $Z_b$ computed from the unit cell ABCD matrix with the Bloch impedance as its own termination. (b) The transmission $1-|\Gamma|^2$ through a B2-ATL prototype (blue) presents a low-pass behavior comparable to that of a B2 filter (orange), enabling filter synthesis techniques for ATL design.}
\label{fig:LPFATL}
\end{figure}

Both filters and ATLs can be characterized by similar concepts. A filter is characterized by its input impedance $Z_1$ looking into port $1$, when port $2$ is terminated by a $1$\,\ohm{} load impedance:
\begin{equation}
    Z_1 = \frac{A+B}{C+D} \label{eq:Z1}.
\end{equation}
Here, $A$, $B$, $C$, and $D$ are the filter ABCD matrix parameters (each being function of $\omega$). In an ATL, the infinite cascade of unit cells can be modeled by taking the input impedance at port $1$ of one cell, with its port $2$ terminated by the input impedance itself \cite{caloz2005electromagnetic}, see Fig.\,\ref{fig:LPFATL}a. This defines the ATL Bloch impedance:
\begin{equation}
    Z_b = \frac{A Z_b+B}{C Z_b+D} \label{eq:Z1ATL}.
\end{equation}
Solving the quadratic equation in $Z_b$ when $C\neq0$ yields:
\begin{equation}
    Z_\mathrm{b} = \frac{A-D}{2C} \pm \frac{j}{C}\sqrt{1-\left(\frac{A+D}{2}\right)^2}.
\label{eq:Zb}
\end{equation}
For a lossless filter, $\{A,D\}\in\mathbb{R}$ while $\{B,C\}\in i\mathbb{R}$, therefore the first term in $Z_b$ is always purely imaginary. The second term can vary as a function of frequency: it is either purely real (for a positive radicand), in which case the associated mode is propagating, or purely imaginary (negative radicand), in which case the associated mode does not propagate. When the unit cell is symmetric, $A=D$ (while $B=C$ for an antimetric unit cell). The reciprocity condition $AD-BC=1$ then simplifies Eq.\,\ref{eq:Zb} to $Z_b = \pm\sqrt{B/C}$, where the sign distinguishes forward ($\Re{Z_b}>0$, group velocity $v_g>0$) from backward ($\Re{Z_b}<0$, $v_g<0$) propagating waves. In Eq.\,\ref{eq:Zb}, forward propagation corresponds to selecting the $\pm$ sign to match $\operatorname{sgn}(j/C)=\operatorname{sgn}(\Im{C})$.

Excluding any spatially periodic modulation of the series and shunt elements along the ATL, the only permitted unit cell prototype corresponds to a second order ($L-C$) or third order ($L-C-L$) LPF. To minimize the number of components per unit cell, I focus on the second order (but note that the synthesis can be carried out with a third order LPF in a similar fashion). In this context, a Butterworth unit cell, where $L=C$ (in normalized units), represents a regular right-handed ATL. It is perfectly impedance matched at $\omega=0$ and maximally flat for all frequencies well below the line's cutoff frequency. More precisely, the outcome of the filter synthesis procedure is an array of numbers $\{g\}$ representing the LPF normalized component values (including the source and load impedance), see appendix \ref{app:filter_synth}. A second-order Butterworth LPF (B2-LPF) is characterized by $g=\{1,\sqrt{2},\sqrt{2},1\}$, meaning that the normalized inductor and capacitor values are $g_L=\sqrt{2}$ and $g_C=\sqrt{2}$, respectively. In other words, its ABCD matrix is
\begin{equation}
\begin{aligned}
    \mathrm{ABCD}_\mathrm{B2} &= \begin{pmatrix} 1 & s\sqrt{2} \\ 0 & 1 \end{pmatrix}\begin{pmatrix} 1 & 0 \\ s\sqrt{2} & 1 \end{pmatrix} \\
    &= \begin{pmatrix} 1+2s^2 & s\sqrt{2} \\ s\sqrt{2} & 1 \end{pmatrix},
\end{aligned}
\label{eq:ABCDB3}
\end{equation}
where $s=\sigma+j\omega$ is the complex frequency. When $\lvert s=j\omega\rvert\ll1$ (far from the ATL's cutoff frequency), $Z_b=\pm1$, \ie{}, the ATL has a constant, real impedance.

Usually, a filter is characterized by its amplitude response $1-\lvert \Gamma \rvert ^2$, where $\Gamma$ is the reflection coefficient off of port $1$, related to the input impedance by
\begin{equation}
    \Gamma(s)=\frac{Z_1(s)-1}{Z_1(s)+1},
\label{eq:gamma}   
\end{equation}
(taking the source impedance to $1$\,\ohm{}). Note that, in general, the zeros (roots of the numerator) and poles (roots of the denominator) of $\Gamma$ lie in the complex plane, see appendix \ref{app:filter_synth}. Similarly, $\Gamma$ can be defined for an ATL by replacing $Z_1$ with $Z_b$ in Eq.\,\ref{eq:gamma}, and then $1-\lvert \Gamma \rvert^2$ quantifies how well the semi-infinite ATL is matched to a unitary source impedance, see Fig.\,\ref{fig:LPFATL}a.

Figure \ref{fig:LPFATL}b presents the transmission $1-\lvert \Gamma \rvert^2$ through a B2-LPF, and through a semi-infinite B2-ATL prototype. Both filter and ATL have perfect transmission ($\lvert\Gamma\rvert^2=0$) at the same frequencies, see appendix \ref{app:zeropolesATL}. As expected from Eqs.\,\ref{eq:Zb} and \ref{eq:ABCDB3}, the transmission through the B2-ATL is unitary for $\omega\ll1$, decreases as $\omega\rightarrow\sqrt{2}$, and is null above the cutoff frequency $\omega_c=\sqrt{2}$, where $Z_b\in i\mathbb{R}$.
    
\subsubsection{Filter transformations: engineering any pass-stop response}
\label{sec:freq_trans}

The low-pass ATL prototype thus supports a continuum of propagating modes, from $\omega=0$ to $\omega\rightarrow\sqrt{2}$. As in filter synthesis, frequency transformations $s\rightarrow\lambda(s)$ can selectively control which modes propagate, by mapping the low-pass ATL response at $s=0$ and $s\rightarrow\infty$ to frequencies where $\lambda=0$ and $\lambda\rightarrow\infty$, respectively (here $s=j\omega$). For a filter, this transformation can be written as a reactance function \cite{foster1924a,guillemin1957synthesis}:
\begin{equation}
    \lambda(s)=Hs\frac{(s^2+\omega_1^2)...(s^2+\omega_{2n+1}^2)}{(s^2+\omega_0^2)...(s^2+\omega_{2n}^2)},
\label{eq:reaczero}
\end{equation}
or
\begin{equation}
    \lambda(s)=\frac{H}{s}\frac{(s^2+\omega_1^2)...(s^2+\omega_{2n-1}^2)}{(s^2+\omega_2^2)...(s^2+\omega_{2n}^2)},
\label{eq:reacpole}
\end{equation}
where $H$ is a scale factor, and where the degrees of the numerator and denominator must only differ by $\pm1$, because at $\omega=0$ and $\omega\rightarrow\infty$ the reactance should behave either as an inductor, or as a capacitor \cite{foster1924a}. Equation \ref{eq:reaczero} (Eq.\,\ref{eq:reacpole}) corresponds to the case where $\lambda$ has a zero (pole) at $\omega=0$ (at $\omega\rightarrow\infty$ the behavior is determined by the number of poles and zeros). Importantly, such a reactance can be decomposed into partial fractions as:
\begin{equation}
    \lambda(s) = \frac{k_0}{s} + \sum_ {i}\frac{2k_{2i} s}{s^2+\omega_{2i}^2} + k_\infty s,
\label{eq:lambda}    
\end{equation}
where $\{k_0,k_{2i},k_\infty\}$ are the residues of $\lambda$ in its poles at $s=0$, $s=\pm j\omega_{2i}$ and $s=\infty$, respectively, if they exist (if a pole does not exist, the corresponding residue is zero). These residues can be calculated using \cite{guillemin1957synthesis}:
\begin{align}
    k_0 &= \left.s\lambda(s)\right\rvert_{s=0}\\
    k_{2i} &= \left.(s-j\omega_{2i})\lambda(s)\right\rvert_{s=j\omega_{2i}}\\
    k_\infty &= \left.\frac{\lambda(s)}{s}\right\rvert_{s\rightarrow\infty}.
\end{align}

Equivalently, the same transformation can be written as $s^{-1}\rightarrow\lambda^{-1}(s)$. Here, $\lambda^{-1}$ is a susceptance function, with zeros and poles being the poles and zeros of $\lambda$, respectively. It can also be expanded into partial fractions, similarly to Eq.\,\ref{eq:lambda}, albeit with different residues. These two possible network transformations using $\lambda$ or $\lambda^{-1}$, are called the \textit{Foster} forms $1$ and $2$, respectively \cite{foster1924a,guillemin1957synthesis}.

\begin{figure*}[hptb!]
\includegraphics[width=\textwidth]{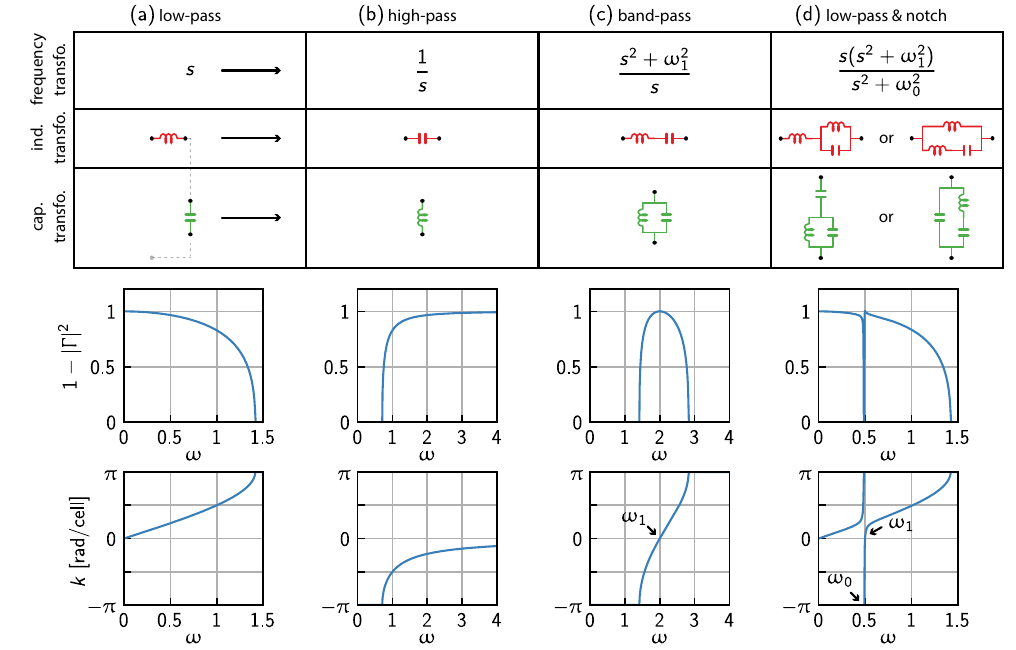}
\caption{Filter transformations of a low-pass ATL prototype. (a) Starting from a low-pass ATL whose unit cell comprises a series inductance and a shunt capacitance, systematic frequency transformations of the complex frequency $s=j\omega$ create different pass-stop responses. (b) The high-pass transformation inverts the component types, creating a left-handed ATL with $k < 0$ in the passband. (c) The band-pass transformation produces series and parallel LC resonators, yielding an ambidextrous (composite right-left-handed) ATL with the passband transitioning from $k < 0$ to $k > 0$ at the transmission zero $\omega_1$. (d) The low-pass and notch transformation creates either series or parallel filter configurations, introducing a transmission pole at $\omega_0$, and a zero at $\omega_1$. For each transformation, the transmission $1 - |\Gamma|^2$ and wavenumber $k$ demonstrate the distinct amplitude and phase characteristics of the resulting ATL.}
\label{fig:transfo}
\end{figure*}

In terms of circuits, the expansion of the reactance, Eq.\,\ref{eq:lambda}, corresponds to a capacitor (first term) in series with parallel LC resonators (second term), and with an inductor (third term). Equivalently, the expansion of the susceptance $\lambda^{-1}$ would correspond to an inductor, in parallel with series LC resonators, and with a capacitor. So, for each component in the LPF prototype there exist two possible transformations: each inductor (capacitor) can be viewed as an impedance (admittance) $g_ks$ which transform into $g_k\lambda(s)$, or as an admittance (impedance) $g_k^{-1}s^{-1}$ which transforms into $g_k^{-1}\lambda^{-1}(s)$. In general, the two different partial fraction expansions of $\lambda$ and $\lambda^{-1}$ lead to two different sets of component values and, in practice, usually one of the sets of values corresponds to components easier to obtain or fabricate. For the simple high-pass and band-pass cases, these two transformations degenerate into one. Appendix \ref{app:filterfunc} shows functions, used to calculate the transformed filter coefficients for each Foster form.

Such frequency transformations are not specific to a particular prototype circuit, therefore an ATL prototype can be transformed following the same procedure. Figure \ref{fig:transfo} shows frequency transformation examples, and the associated B2-ATL amplitude response $1-\lvert\Gamma\rvert^2$, along with the phase response, characterized by the wavenumber $k$, calculated for $v_g>0$ from the unit cell ABCD matrix using:
\begin{equation}
    k = \operatorname{sgn}(\Im{C})\Im{\arccosh\left(\frac{A+D}{2}\right)},
\label{eq:k}
\end{equation}
see appendix \ref{app:kgen}. The high-pass ATL (Fig.\,\ref{fig:transfo}b), also called \textit{left-handed} \cite{caloz2005electromagnetic}, has the remarkable property of having $k<0$ in its passband. In other words, the phase velocity $v=\omega/k$ of the high-pass ATL is negative, while $v_g=\dd{\omega}/\dd{k}>0$, because $k(\omega)$ has a positive slope. The band-pass ATL (Fig.\,\ref{fig:transfo}c) presents a composite left-handed ($k<0$, $v_g>0$) and right-handed ($k>0$, $v_g>0$) behavior in its passband, which I will use in Sec.\,\ref{sec:bTWPA} to create a backward-pumped TWPA.

\subsubsection{Implementation flexibility}

The low-pass and notch ATL (Fig.\,\ref{fig:transfo}d) corresponds to a typical TWPA design that uses a resonant phase-matching (rpm) filter. There are two filter topologies, each for the series and shunt branch of the unit cell. But the ATL achieves the same response without filtering both branches. At the non-trivial transmission zero $\omega_1$, the series filter is a short while the shunt filter is an open, so that all the signal is perfectly transmitted along the line. The signal is also perfectly transmitted if the series filter is not a perfect short, so long as the shunt filter is a perfect open, and vice versa. Conversely, at the transmission pole $\omega_0$, the signal is perfectly reflected with either a series open or a shunt short, not necessarily both. Compared to standard passive filters that are matched to fixed input and output real impedances, this additional degree of freedom in ATL arises because each cell is inherently matched to its neighbors, regardless of the filter configuration.

This flexibility allows implementing the rpm filter with any of the four configurations in Fig.\ref{fig:transfo}d. JTWPAs typically use the shunt capacitor in series with the parallel LC resonator \cite{macklin2015near,white2015traveling,qiu2023broadband,malnou2025a}, because the desired low mode impedance $\sqrt{L/C}$ suits multi-layer fabrication techniques to create large capacitors. For single-layer KTWPAs, the series inductor in parallel with a series LC resonator is preferable, because the desired high mode impedance suits fabrication with high kinetic inductance materials.

The ATL also approximately achieves the same response without filtering every unit cell, despite creating impedance mismatches at the filtered/unfiltered cell boundaries. While even a single series open or shunt short suffices to create the transmission pole \ie{}, perfect reflection, the transmission zero degrades when filters are spaced beyond a quarter wavelength at $\omega_1$, disrupting the impedance matching required for perfect transmission.

\subsubsection{Band structure and Foster's reactance theorem in k-space}
\label{sec:kfoster}

In each example of Fig.\,\ref{fig:transfo}, $k$ is always a strictly monotonically increasing function of $\omega$, although not necessarily continuous. Furthermore, $v_g=0$ only at the edge of the first Brillouin zone \ie{}, for $k=\pm\pi$. This is a consequence of the lossless nature of the ATL, and can be viewed as an extension of Foster's reactance theorem \cite{foster1924a} to $k$-space, see appendix \ref{app:kFoster}. Note that it is true for any lossless ATL, including that designed via periodic loading.

Consequently, when creating a passband via frequency transformation (Fig.\,\ref{fig:transfo}c), $k$ increases monotonically from $-\pi$ to $\pi$, passing through $k=0$ where $v_g>0$. In other words, a passband ATL presents both a left- and a right-handed behavior, corresponding to two separate frequency ranges.

\subsection{Comparison between both approaches and limits to the synthesis}

\subsubsection{Topological considerations}
\label{sec:topo_considerations}

It may appear that a periodically loaded ATL has alternating right and left-handed bands, see Fig.\,\ref{fig:plsynth}. In fact, within the  irreducible Brillouin zone (IBZ), where $k\in[0,\pi/d]$, bands where $v_g>0$ and $v_g<0$ alternate with increasing frequency. However, the bands where $v_g<0$ are not left-handed, because the wavenumber $k$ to consider is that of the mode creating that band, not any equivalent wavenumber that appears in the IBZ due to periodicity. The first passband comes from propagation of the fundamental spatial harmonic ($n=0$), the second one comes from propagation of the second spatial harmonic ($n=-1$) and so on. More formally, what matters is $k_B + n k_d$, where $k_B$ is the Bloch wavenumber (\ie{}, $k$ restricted to the first Brillouin zone), and $n$ corresponds to the dominant spatial harmonic for the particular band under consideration. For example, within the IBZ the second passband has $v_g<0$, but $k-k_d<0$ also, therefore the line remains right-handed. In other words, a periodically loaded ATL with low-pass unit cells is always right-handed because the loading does not affect its \textit{topology}. In particular, it will never have a stopband at dc. Conversely, the filter synthesis approach can truly introduce left-handedness, because the filter transformations affect the propagation of the first spatial harmonic.

Both the periodic loading and the filter synthesis can be accomplished within the same ATL. In fact, the filter transformations presented in Sec.\,\ref{sec:freq_trans} can also be applied to a periodically loaded ATL, in which case the response from $\omega=0$ to $\omega=\infty$, that includes possible stopbands, will be mapped to the intervals between which $\lambda(\omega)=0$ and $\lambda(\omega)=\infty$ (or $\lambda^{-1}(\omega)=\infty$ and $\lambda^{-1}(\omega)=0$), where $\lambda$ ($\lambda^{-1})$ is the reactance (susceptance) function describing the transformation.

Moreover, a finite ATL may contain aperiodic modulations of its components: it is for example the case in a Floquet TWPA \cite{pengPRXQuantumFloquet,wang2025high}, or in a periodically-loaded TWPA with apodized inputs and outputs \cite{chang2025josephson}. In that case, the spatial harmonic basis $\{e^{-j n k_d x}\}_{n\in\mathbb{Z}}$ used to describe the ATL is inconvenient, because then the entire finite ATL constitutes the supercell.

\subsubsection{Practical implementations}

In practice, the phase-matching condition in TWPAs can be achieved via periodic loading \cite{Eom2012a,vissers2016low,malnou2021three,chang2025josephson,howe2025kinetic}, but it presents two issues: (i) in a \textit{finite} ATL, it creates ripples near the stopband \cite{gaydamachenko2022numerical}, where the optimal pump frequency for phase-matching usually is. Thus, the phase-matched pump may suffer from an unfavorable impedance mismatch, which leads to pump reflections and backward gain. Furthermore, these ripples may land within the amplification band, affecting the gain profile. Apodization techniques can mitigate these spurious effects \cite{chang2025josephson}. (ii) When generated from periodic loading, the phase shift near a stopband only slowly approaches the edge of the first Brillouin zone $\pm\pi/d$, with $d$ the number of cells per supercell. This phase shift is limited by the loading value. In contrast, filter synthesis can achieve a sharp jump to the same $\pm\pi/d$ limit, but the transition sharpness is controlled by the pole-zero separation\cor{, which in practice is constrained by achievable component values,} rather than spatial harmonic coupling (see Fig.\,\ref{fig:KTWPA_DE}c). Consequently, achieving phase-matching via periodic loading requires heavily loading the ATL, creating large stopband regions unusable for amplification, while filter synthesis provides localized phase shift, with \cor{a bandwidth penalty set by the pole-zero separation}.

Furthermore, TWPAs leveraging rpm filters usually do not have such filters in every unit cell. Instead, only a few unit cells are periodically modified \cite{macklin2015near,white2015traveling,qiu2023broadband,malnou2025a}. While the resulting supercell can be viewed as a multi-pole filter, the presented periodic loading and filter syntheses do not provide a systematic method to optimally place its zeros and poles in the complex frequency plane, in order to achieve a desired ATL response. However, as shown in Sec.\,\ref{sec:4WMKTWPA}, the rpm filters can still be designed via the ATL filter synthesis theory, even when these are not placed in every unit cell, because far from its cutoff frequency the response of the ATL is unaffected by this non-uniformity.

\section{ATL synthesis for traveling-wave parametric amplifiers}
\label{sec:examples}

I now adopt a semi-systematic approach to design two TWPAs: a kinetic inductance-based device (KTWPA) operating in four-wave mixing (4WM), and a Josephson junction-based device (JTWPA) operating in three-wave mixing (3WM). The approach is semi-systematic rather than fully systematic because the ATL parameters are not directly optimized from a target nonlinear gain response. Instead, I first leverage the presented ATL synthesis to engineer the linear dispersion relation, ensuring the desired phase-matching conditions for parametric amplification. Subsequently, I introduce nonlinearity by incorporating either kinetic inductance elements or Josephson junctions within each unit cell, placing them in the series sections, where the current predominantly flows to achieve efficient parametric mixing. The nonlinear response is then calculated using JosephsonCircuits.jl \cite{pengPRXQuantumFloquet,levochkina2025Modeling,wang2025high}, an open-source harmonic balance simulator that I have extended to support arbitrary nonlinear inductances beyond Josephson junctions, defined through their inductance-flux Taylor expansion \cite{malnou2025twpa}. This separation of linear and nonlinear design is particularly advantageous for 3WM processes, where the phase-matching condition remains unchanged by the nonlinearity, unlike 4WM where self- and cross-phase modulation modify the dispersion.

\subsection{A four-wave mixing KTWPA}
\label{sec:4WMKTWPA}

A KTWPA relies on the nonlinearity of the kinetic inductance. For a $n$ squares-long wire made of kinetic inductance material \cite{Zmuidzinas2012superconducting,Eom2012a}:
\begin{equation}
    L = L_0\left[1+\left(\frac{I}{I_*}\right)^2\right],
\label{eq:LKI}
\end{equation}
where $L_0=nL_\square$ with $L_\square$ the sheet inductance, $I$ is the current flowing through the wire, and $I_*$ is the scaling current, on the order of (but usually higher than) the wire's critical current \cite{howe2025kinetic}. The scale of the nonlinearity is thus inversely proportional to $I_*$, which typically ranges from several milliamperes down to a few hundreds of microamperes. In comparison, a JTWPA's nonlinearity is stronger, as it is set by the junctions' critical current, typically on the order of a microampere. Consequently, KTWPAs often require more unit cells and higher pump power to achieve comparable gain.

\subsubsection{Design choices}

I design the KTWPA with $50$\,\ohm{} characteristic impedance, and choose $I_* = 100$\,\uA{} and $L_0=100$\,pH, values that are practically achievable \cor{\cite{malnou2021three,howe2025kinetic,howe2025compact,larson2025localized}} and suitable for the readout of tens of frequency multiplexed qubits. To get phase-matching for 4WM amplification, I synthesize a low-pass and notch ATL with a series LC rpm filter (see Fig.\,\ref{fig:transfo}d), creating a non-trivial pole-zero pair. The pole is positioned slightly above the desired phase-matched pump frequency $f_p$, so that it can affect the pump wavenumber $k_p$ to compensate for the cross-phase modulation being stronger than the self-phase modulation. The zero is positioned as close to the pole as possible to minimize the bandwidth of the notch in transmission, while keeping the filter component values within reasonable fabrication limits.

In addition, the propagation of the pump's third harmonic (detrimental to traveling-wave parametric amplification \cite{Landauer1960shock}) is suppressed by placing a large stopband around $3f_p$, engineered via periodic modulation. While third harmonic suppression is often overlooked in JTWPAs due to their stronger chromatic dispersion (arising from the junction's shunt capacitance that lowers the line's cutoff frequency), it is essential in KTWPAs: their weaker dispersion means the third harmonic is more closely phase-matched and can therefore propagate efficiently if not purposefully suppressed.

\begin{figure}[h!]
\includegraphics[width=\columnwidth]{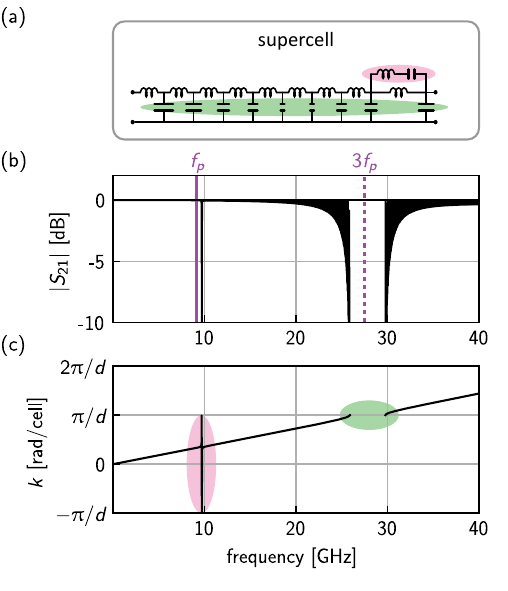}
\caption{Linear response design of the 4WM KTWPA. (a) The supercell comprises a $9$-cell LC ladder with periodically modulated shunt capacitors, where one unit cell contains a series LC resonator (rpm filter) placed in parallel with the series inductor. (b) The power transmission $|S_{21}|$ (in decibel) through a $5004$-cell ATL shows the phase-matching notch at $10$\,GHz created by the rpm filter, and the third-harmonic stopband at $27$\,GHz, created by the periodic capacitor modulation. (c) The wavenumber $k$ of the corresponding infinite lattice demonstrates rapid phase accumulation near $10$\,GHz essential for 4WM phase matching, and the bandgap at $27$\,GHz that suppresses third-harmonic generation.}
\label{fig:KTWPA_DE}
\end{figure}

\subsubsection{Simulation results}

\begin{figure}[h!]
\includegraphics[width=\columnwidth]{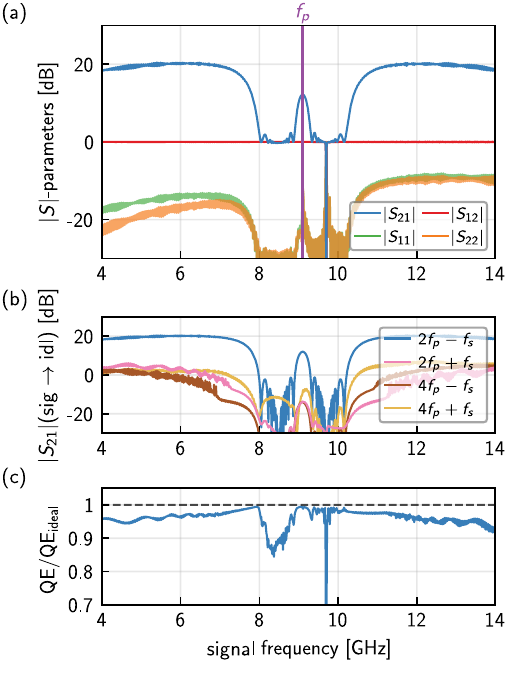}
\caption{Simulated nonlinear response of the 4WM KTWPA. (a) The full 2-port scattering parameters as a function of frequency show the TWPA's flat $20$\,dB gain profile from $4$ to $8$\,GHz with matched inputs and outputs ($\lvert S_{11}\rvert, \lvert S_{22}\rvert < -10$\,dB). (b) Forward idler tone analysis demonstrates effective third-harmonic suppression, with the first idler at $2f_p - f_s$ amplified as expected while higher-order mixing products near $3f_p$ are suppressed by $\sim20$\,dB due to the engineered stopband. (c) Quantum efficiency exceeds $95$\% of the theoretical maximum across the full $4$-$8$\,GHz bandwidth, confirming clean 4WM parametric conversion with minimal spurious process degradation.}
\label{fig:KTWPA_sim}
\end{figure}

Figure \ref{fig:KTWPA_DE} shows the linear response of a KTWPA. The rpm filter within the supercell (Fig.\,\ref{fig:KTWPA_DE}a) creates a sharp notch at $10$\,GHz, and the periodically modulated shunt capacitor creates a wide stopband centered around $27$\,GHz. Both features appear on the calculated power transmission (Fig.\,\ref{fig:KTWPA_DE}b) through the \textit{finite} ATL of $5004$ unit cells ($556$ supercells). The wavenumber $k$ of the corresponding \textit{infinite} lattice (Fig.\,\ref{fig:KTWPA_DE}c, calculated using Eq.\,\ref{eq:k}) shows the rapid phase accumulation near $10$\,GHz needed for phase matching, as well as the bandgap around $27$\,GHz. The design achieves these responses with practical component values: maximum inductance of 4.9 nH and capacitance of 54 fF (both in the LC filter), with detailed values provided in appendix \ref{app:values}.

I simulate the nonlinear response of the KTWPA using the extended version of JosephsonCircuits.jl \cite{pengPRXQuantumFloquet,levochkina2025Modeling,malnou2025twpa} supporting arbitrary inductance-phase Taylor nonlinearities. For the kinetic inductance nonlinearity, Eq.\,\ref{eq:LKI}, it corresponds to 
\begin{equation}
    L = L_0\left[1+\left(\frac{I_0}{I_*}\varphi\right)^2\right],
\label{eq:Lphi}
\end{equation}
where $I_0=\varphi_0/L_0$ is the inductor's characteristic current (distinct from its critical current), $\varphi_0$ is the reduced magnetic flux quantum, and where $\varphi$ is the reduced branch phase across the inductor (see appendix \ref{app:Lphi}).

Pumping the KTWPA at the phase-matched frequency ($f_p=9.1$\,GHz) with amplitude $I_p=22$\,\uA{} yields the performance shown in Fig.\,\ref{fig:KTWPA_sim}. The device achieves $20$\,dB flat forward gain from $4$ to $8$ GHz, with unity backward gain (Fig.\,\ref{fig:KTWPA_sim}a). Impedance matching remains excellent with $\lvert S_{11}\rvert$ and $\lvert S_{22}\rvert$ below $-10$\,dB. The forward idler response (Fig.\,\ref{fig:KTWPA_sim}b) demonstrates the effectiveness of the third-harmonic stopband: while the first idler at $2f_p-f_s$ (with $f_s$ the signal frequency) is amplified, the second and third idlers near $3f_p$ are suppressed by $\sim20$\,dB \cor{relative to the first idler at $2f_p - f_s$ (appendix \ref{app:thirdharmonic} further illustrates the role of this stopband in enforcing 4WM amplification)}. Concurrently, the quantum efficiency (Fig.\,\ref{fig:KTWPA_sim}c) maintains $>95\%$ of its theoretical maximum across the full $4$-$8$\,GHz bandwidth.

\subsection{An ambidextrous TWPA}
\label{sec:bTWPA}

In a TWPA built from a broadband ATL, unwanted parametric processes can propagate \cite{vissers2016low,erickson2017theory,Zorin2021quasi,remm2023intermodulation,malnou2025a}. These processes deplete pump photons and degrade the TWPA noise performance \cite{pengPRXQuantumFloquet}. The KTWPA presented in Sec.\,\ref{sec:4WMKTWPA} used a stopband in the otherwise broadband ATL transmission to suppress third-harmonic generation. Here, I pursue an alternative strategy: synthesizing a band-pass ATL (Fig.\,\ref{fig:transfo}c) to intrinsically limit the propagation bandwidth. In that situation, within its passband the line is left-handed for frequencies $\omega<\omega_1$ (with $\omega_1$ the transmission zero) and right-handed for $\omega>\omega_1$. Right-handed TWPAs dominate the experimental landscape in superconducting circuits \cite{Eom2012a,macklin2015near,planat2020photonic,malnou2021three,gaydamachenko2025rf,chang2025josephson}. Although left-handed metamaterials are well-established in optics \cite{popov2006compensating,popov2009resonant,suchowski2013phase,lan2015backward}, at microwave frequencies only preliminary demonstrations in non-superconducting platforms exist for parametric amplification \cite{kozyrev2006parametric} and second-harmonic generation \cite{rose2011controlling}. While left-handed superconducting TWPAs have been proposed theoretically \cite{kow2025self}, the synthesis framework reveals a third path: constructing a superconducting TWPA from an ambidextrous transmission line \ie{}, composite right-left-handed \cite{caloz2005electromagnetic}. Such a device offers several advantages: the pump harmonics are strongly mismatched or suppressed due to their proximity to the line's cutoff frequency, and the low-frequency noise \cite{malnou2025a} is suppressed due to the high-pass characteristic.

\subsubsection{Backward phase-matching}

In a 3WM TWPA the phase-matching condition follows directly from the linear dispersion relation: $k_p = k_s + k_i$, where $k_p$, $k_s$, and $k_i$ are the wavenumbers of the pump, signal and idler frequencies, respectively. Unlike 4WM, which requires nonlinear corrections, it allows direct use of the band diagram from the ATL synthesis framework to directly identify where and how this wavenumber relationship can be satisfied. Additionally, the pump frequency naturally falls above the amplification band, \cor{significantly relaxing pump filtering requirements}.

\begin{figure}[h!]
\includegraphics[width=\columnwidth]{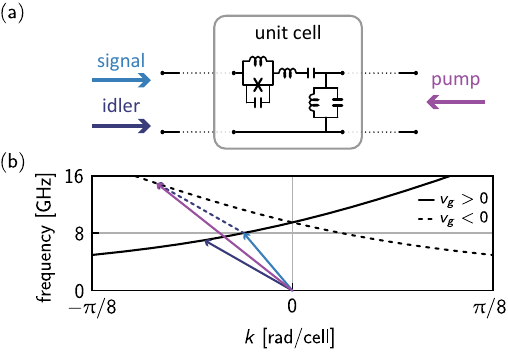}
\caption{Backward phase-matching in a b-TWPA. (a) The unit cell comprises a series LC resonator (series branch) with an rf-SQUID providing nonlinearity, and a parallel LC resonator (shunt branch). The signal and idler propagate forward while the pump propagates backward to achieve phase matching in the ambidextrous ATL. (b) In the band diagram, the vector sum of signal (blue) plus idler (dark dashed blue), both in the left-handed region, equals the backward-propagating pump vector (purple) in the right-handed region, satisfying $f_p = f_s + f_i$ and $k_p = k_s + k_i$ for 3WM parametric amplification.}
\label{fig:bTWPA_PM}
\end{figure}

Visually, respecting the 3WM phase-matching condition and the conservation of energy corresponds to the pump, signal, and idler vectors forming a closed loop in the band diagram. Starting from signal and idler frequencies in the left-handed portion of the ATL so that $f_p = f_s + f_i$ still falls within the passband (Fig.\,\ref{fig:bTWPA_PM}b), the vector sum $k_s + k_i$ does not intersect the dispersion relation, precluding phase matching, unless the pump is allowed to propagate backward, residing on the branch where $v_g < 0$. Remarkably, in 4WM this ambidextrous ATL supports another exotic phase-matching scheme, where the signal or idler back-propagates while the other forward-propagates with the pump. This configuration enables the generation of entangled counter-propagating photon pairs and mirrorless resonances \cite{popov2009resonant,popov2006compensating,praquin2024mixing}.

\subsubsection{The b-TWPA unit-cell}

The ambidextrous ATL unit cell comprises a series LC resonator in the series branch and a parallel LC resonator in the shunt branch (Fig.\ref{fig:transfo}c). Component values are determined by two key parameters: the transmission zero frequency $f_1$ and the ATL cutoff frequency $f_c$. To center the amplification band around $7.5$\,GHz, the transmission zero is placed at $f_1 = 9.5$\,GHz. The cutoff frequency selection involves a critical trade-off: higher values yield flatter dispersion and broader amplification bandwidth, but require larger component values and permit unwanted parametric processes. Choosing $f_c = 48$\,GHz ensures that the pump's second harmonic ($\sim30$\,GHz) experiences significant attenuation near cutoff while maintaining practical component values: maximum inductance of $3$\,nH (shunt branch) and capacitance of $1.2$\,pF (series branch), both achievable with standard multi-layer superconducting circuit fabrication \cite{lecocq2017nonreciprocal,malnou2025a}.

To enable 3WM parametric amplification, an rf-SQUID replaces part of the linear inductor in the series branch of each unit cell, see Fig.\,\ref{fig:bTWPA_PM}a. The junction's critical current $I_c = 2$\,\micro A preserves sufficient power handling for qubit readout applications, and a screening parameter $\beta_L = 0.4$ is suitable for stable rf-SQUID amplifier operation \cite{kaufman2025simple,gaydamachenko2025rf}. The rf-SQUIDs are flux-biased at their Kerr-free point, where $\varphi_\text{dc} = \pi/2$ (corresponding to $\Phi_\text{ext}/\varphi_0 = \varphi_\text{dc} + \beta_L \sin{\varphi_\text{dc}} \approx 2$, with $\varphi_0$ the reduced flux quantum), enabling 3WM while suppressing spurious 4WM processes. At this operating point, the rf-SQUID static inductance is $L_g = \beta_L \varphi_0/I_c \simeq 66$\,pH. Each junction introduces a parasitic capacitance from its $\sim40$\,GHz plasma frequency. Though not included in the initial ATL synthesis, this capacitance only slightly perturbs the linear response, creating a narrow stopband near the transmission zero (see Fig.\,\ref{fig:bTWPA_sim}a), but does not significantly affect the phase-matching condition.

\subsubsection{Simulation results}

\begin{figure}[h!]
\includegraphics[width=\columnwidth]{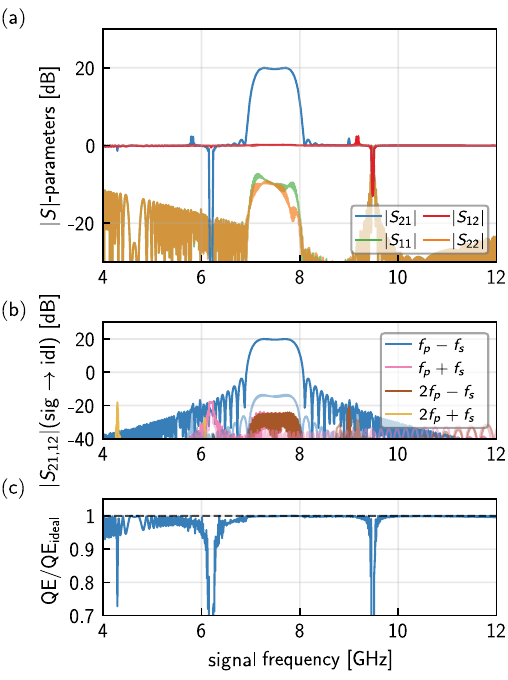}
\caption{Simulated nonlinear response of the b-TWPA. (a) The full 2-port scattering parameters as a function of frequency show the TWPA's wideband $20$\,dB gain profile. The dips in $\lvert S_{21} \rvert$ at $9.5$\,GHz and $6.25$\,GHz correspond to the transmission zero and its corresponding idler, respectively. (b) Forward (plain lines) and backward (faint lines) idler tone analysis reveals effective suppression of unwanted parametric processes, with higher-order mixing products below $-20$\,dB while the first idler at $f_p - f_s$ is amplified as expected. The faint blue trace shows the backward-propagating first idler, suppressed by over $30$\,dB relative to its forward counterpart. (c) Quantum efficiency approaches the theoretical maximum across the amplification bandwidth, demonstrating near-perfect parametric conversion.}
\label{fig:bTWPA_sim}
\end{figure}

Figure \ref{fig:bTWPA_sim} presents the nonlinear response of a $2000$-unit-cell backward-pumped TWPA (b-TWPA), simulated with the extended version of JosephsonCircuits.jl \cite{malnou2025twpa}. The biased rf-SQUIDs are modeled via their Taylor expansion equivalent. Using a pump frequency $f_p = 14.98$\,GHz with amplitude $I_p = 1.1$\,\micro A, the $\lvert S\rvert$-parameters (Fig.\ref{fig:bTWPA_sim}a) show a forward gain $\lvert S_{21} \rvert=20$\,dB across $\sim1$\,GHz of bandwidth centered around $7.5$\,GHz. Meanwhile, the backward gain $\lvert S_{12}\rvert$ remains near $0$\,dB, confirming directional amplification. The device maintains good impedance matching with return losses ($\lvert S_{11}\rvert$  and $\lvert S_{22} \rvert$) generally below $-10$\,dB across the amplification band, despite a modest degradation in $\lvert S_{11} \rvert$ near the lower band edge. 

The idler response (Fig.\ref{fig:bTWPA_sim}b) demonstrates the effective suppression of unwanted parametric processes. While the first idler at $f_p - f_s$ is amplified alongside the signal as expected for phase-insensitive parametric amplification, all higher-order mixing products ($f_p + f_s$, $2f_p - f_s$, and $2f_p + f_s$) are effectively suppressed, remaining below $-20$\,dB. Notably, the backward-propagating first idler exhibits slightly elevated levels compared to the higher-order products, yet remains more than $30$\,dB below its forward-propagating counterpart, further confirming the device's directional operation. This suppression of spurious parametric processes enables a near-perfect quantum efficiency (Fig.\ref{fig:bTWPA_sim}c) across the entire amplification bandwidth.

\section{Conclusion}

\cor{I have developed a theoretical framework for designing artificial transmission lines, encompassing two complementary synthesis approaches}: periodic loading synthesis, where frequency-independent components are spatially modulated to create stopbands in right-handed ATLs, and filter synthesis, where spatially-invariant components are engineered with frequency-dependent responses. Periodic loading synthesis relies on \cor{a generalized matrix dispersion relation that can be used for stopband design}, found by decomposing voltage, current, and component modulations onto a common spatial harmonic basis. \cor{ATL design via filter synthesis recognizes that a right-handed ATL behaves as a low-pass prototype, enabling direct application of filter transformations}.

\cor{This framework} reveals fundamental constraints on ATL dispersion relations: \cor{in particular,} the wavenumber $k$ is always a monotonically increasing function of frequency, though not necessarily continuous, a result that parallels Foster's reactance theorem in the context of artificial transmission lines. \cor{It also reveals} previously unexplored parametric amplifier architectures.

I validated this approach by designing and simulating two novel TWPAs using an extended open-source harmonic balance solver that supports arbitrary nonlinear inductances defined through their inductance-flux Taylor expansion. The first, a 4WM KTWPA, employs a new rpm filter topology particularly suited for kinetic inductance devices. The second demonstrates a new concept: a backward-pumped TWPA (b-TWPA) based on an ambidextrous transmission line where the pump propagates backward while signal and idler propagate forward. 

\cor{Superconducting circuits offer unique advantages for exploring these new concepts: they can be constructed from the ground up using lumped inductors and capacitors while maintaining ultra-low loss. This flexibility, however, generates a vast parameter space. A theoretical framework that constrains this space, such as the one presented here, is therefore a valuable tool for navigating the design process.}

\subsection*{Acknowledgements}

I acknowledge Florent Lecocq, Joe Aumentado, John Teufel, Martin Ritter, Connor Denney, and Cody Scarborough for useful discussions and feedback.

\appendix

\section{Identical periodic modulation of L and C in an ATL}
\label{app:nomodATL}

 The voltage $V(x,t)$ and current $I(x,t)$ at position $x$ and time $t$ along the line are governed by the Telegrapher's propagation equations:
\begin{align}
    \frac{\partial V}{\partial x} &= -L(x) \frac{\partial I}{\partial t} \label{eq:telV}\\
    \frac{\partial I}{\partial x} &= -C(x) \frac{\partial V}{\partial t} \label{eq:telI}.
\end{align}
If the periodic modulation of $L$ is identical to that of $C$, then the ATL is equivalent to an unloaded transmission line, and there are no stopbands. In fact, if $L(x)=L_0 f(x)$ and $C(x)=C_0 f(x)$, where $f$ is a periodic function, then a new space coordinate can be defined
\begin{equation}
    \xi(x) = \int_0^x f(s)ds,
\end{equation}
and in this new coordinate system Eqs.\,\ref{eq:telV} and \ref{eq:telI} are
\begin{align}
    \frac{\partial V}{\partial \xi} &= -L_0 \frac{\partial I}{\partial t}\\
    \frac{\partial I}{\partial \xi} &= -C_0 \frac{\partial V}{\partial t},
\end{align}
which are the propagation equations in an unloaded transmission line. In other words, only the periodic modulation of the ATL's Bloch impedance $Z_b=\sqrt{L/C}$ creates stopbands. Intuitively, without impedance mismatch there can be no reflection, hence no backward waves to which to couple.

\section{Maximum stopband width from single-harmonic modulation}
\label{app:stopwidth}

Considering an ATL whose capacitance contains a single Fourier component, $C(x) = C_0[1 + 2c_1\cos(2\pi x/d)]$, only the $n=0$ and $n=-1$ harmonics couple with each other, and thus Eq.\,\ref{eq:det} reads
\begin{equation}
\begin{vmatrix}
    \tilde{\omega}^2-(\tilde{k}-1)^2 & c_1\tilde{\omega}^2 \\
    c_1\tilde{\omega}^2 & \tilde{\omega}^2-\tilde{k}^2
\end{vmatrix} = 0,
\end{equation}
which yields
\begin{equation}
    (1-c_1^2)\tilde{\omega}^4 - \frac{\tilde{\omega}^2}{2} + \frac{1}{16} = 0,
\end{equation}
when $\tilde{k}=1/2$. The two positive solutions are $\tilde{\omega}_\pm=(2\sqrt{1\pm c_1})^{-1}$, which yields the upper and lower stopband widths
\begin{align}
    \delta_1^- &= \frac{1}{2}\left(1-\frac{1}{\sqrt{1+c_1}}\right)\\
    \delta_1^+ &= \frac{1}{2}\left(\frac{1}{\sqrt{1-c_1}}-1\right).
\end{align}
The stopband's total width relative to $\tilde{\omega}=1/2$ is then $\delta_1=2(\delta_1^++\delta_1^-)$. When $c_1=0.5$ it yields $\delta_1^\mathrm{max}\simeq0.6$. 

\section{The approximation problem in filter synthesis}
\label{app:filter_synth}

In filter synthesis, the approximation problem consists of finding a network of components that approximates a given amplitude or phase response \cite{guillemin1957synthesis}. When synthesizing the amplitude response of a low-pass filter (LPF), the goal is to approximate a frequency step: ideal transmission $\lvert S_{21}(\omega)\rvert^2$ equals unity when $\omega \leq 1$ and zero when $\omega > 1$. The transmission response may be written as
\begin{equation}
    \lvert S_{21}(\omega)\rvert^2 = \frac{1}{1+F_n^2(\omega)},
\label{eq:S21sq}
\end{equation}
where $F_n$ is a real-valued, even or odd polynomial that approximates the desired step response. Common choices include Chebyshev polynomials or $F_n(\omega)=\omega^n$ for Butterworth filters. The polynomial constraint ensures the roots of $F_n^2$ and $1+F_n^2$ occur in quadruplets $\{s_0,s_0^*,-s_0,-s_0^*\}$ where $s = j\omega$, which enables construction of a stable, realizable filter \cite{guillemin1957synthesis}.

Since $\lvert S_{21}\rvert ^2 = 1-\lvert\Gamma\rvert^2$, where $\Gamma$ is the reflection coefficient,
\begin{equation}
    \lvert \Gamma(s)\rvert^2 = \frac{F_n^2(-js)}{1+F_n^2(-js)}.
\label{eq:Gamma2}    
\end{equation}
The relationship $|\Gamma(s)|^2=\Gamma(s)\Gamma^*(s)$ allows construction of the reflection coefficient from its magnitude-squared function: $\Gamma$ is constructed by first selecting the poles of $\lvert\Gamma\rvert^2$ that lie in the left-half plane (\ie{}, with negative real parts), preserving complex conjugate pairs. This ensures filter stability and causality. Note that the poles $s_p$ (where $|\Gamma(s_p)| \rightarrow \infty$) must lie in the complex plane to satisfy the physical constraint that $|\Gamma(\omega)| \leq 1$ for $\omega\in\mathbb{R}$. Conversely, in canonical filter designs the zeros of $\Gamma(s)$ lie on the imaginary axis. For Butterworth filters $F_n(\omega) = \omega^n$, so all zeros are at $s = 0$, while for Chebyshev filters $F_n(\omega) = T_n(\omega)=\cos(n\arccos(\omega))$ so all the zeros are at $s = j\omega$ with $\omega = \cos(\pi(2p+1)/2n)$ for $p \in \{0, 1, \ldots, n-1\}$. Half of the zeros are selected from $\lvert\Gamma(s)\rvert^2$, also preserving complex conjugate pairs, to finally form $\Gamma(s)=N(s)/D(s)$, where $N$ and $D$ are polynomials. This reflection coefficient contains the same number of zeros as poles.

Using Eq.~\ref{eq:gamma}, the input impedance $Z_1$ \cite{naaman2022synthesis} is:
\begin{equation}
    Z_1(s) = \frac{D(s)+N(s)}{D(s)-N(s)}.
\end{equation}
By construction, $\Gamma$ is a bounded-real (BR) function: it has no poles in the right-half plane, satisfies $\Gamma(s^*) = \Gamma^*(s)$ (real coefficients due to conjugate pole-zero pairs), and obeys $|\Gamma(j\omega)| \le 1$. If $\Gamma$ is BR, then $Z_1$ is positive real (PR) \ie{}, $Z_1$ is analytic, real for $s\in\mathbb{R}$ (both properties guaranteed by construction) and $\Re{Z_1(s)}\geq0$. In fact,
\begin{equation}
    \Re\{Z_1(s)\} = \Re\left\{\frac{1+\Gamma}{1-\Gamma}\right\}.
\end{equation}
Multiplying numerator and denominator by $1-\Gamma^*$ gives
\begin{equation}
    \Re\{Z_1(s)\} = 
    \frac{1-|\Gamma|^2}{|1-\Gamma|^2} \ge 0.
\end{equation}
Conversely, starting from a PR impedance $Z_1$ and solving $\Gamma = (Z_1 - 1)/(Z_1 + 1)$ 
yields that $\Gamma$ is BR, so
\begin{equation}
    Z_1 \text{ is PR} \;\Longleftrightarrow\; \Gamma \text{ is BR.}    
\end{equation}
By Darlington’s theorem \cite{guillemin1957synthesis}, $Z_1$ is then realizable as the input impedance of a lossless two-port network terminated in a $1\,\Omega$ load. 
Successive polynomial division of $Z_1(s)$, discarding remainders at each step, yields the normalized filter coefficients $\{g_k\}$ tabulated for classical prototypes such as Butterworth or Chebyshev filters \cite{naaman2022synthesis}.

\section{Relationship between filter and ATL zeros and poles}
\label{app:zeropolesATL}

Here I show that $Z_1=1 \Longleftrightarrow Z_b=\pm1$ \ie{} the reflection zeros of the filter, where $Z_1(s)=1$\,\ohm{}, are the frequencies at which $Z_b(s)=\pm 1$\,\ohm, and reciprocally. If $Z_1=1$ then, from Eq.\,\ref{eq:Z1} $A=D$ and $B=C$, because for a lossless filter $\{A,D\}\in\mathbb{R}$ while $\{B,C\}\in i\mathbb{R}$. Then, when $C\neq0$ using Eq.\,\ref{eq:Zb} with the reciprocal condition $AD-BC=1$ yields $Z_b=\pm1$.

If $Z_b=\pm1$ and with a symmetric unit cell \ie{}, $A=D$, then $Z_b=\pm1=\pm\sqrt{B/C}$ so $B=C$ and therefore $Z_1=1$. If $Z_b=\pm1$ and the unit cell is antimetric \ie{}, $B=C$, then from reciprocity $AD = 1+C^2$. Furthermore 
\begin{equation}
    \pm1 - \frac{A-D}{2C} = \pm\frac{(A+D)^2-4}{2C},
\end{equation}
which leads to $(\pm2C-A+D)^2 = (A+D)^2-4$, which, after simplification, yields $A=D$. So $Z_1=1$.

Note that when $C\rightarrow0$ (for low-pass prototypes, this occurs at $s\rightarrow0$) the equivalence does not hold in general: one can have $Z_1=1$ with $A=D$ (symmetric unit cell), but with $B\sim g_1s$ and $C\sim g_2s$ as $s\rightarrow0$. Therefore $Z_b\rightarrow\sqrt{g_1/g_2}\neq1$. However, the equivalence holds for Butterworth prototypes, where $B\sim C$ as $s\rightarrow0$.

\section{Filter transformation functions}
\label{app:filterfunc}

The frequency transformation of low-pass prototype components uses partial fraction expansion of the reactance or susceptance function. Here I present the formulas for computing transformed component values.

\subsection{Foster form 1 (reactance transformation)}

I consider a reactance function $\lambda(\omega)$ with non-trivial zeros at $\{\omega_{z}\}$ and poles at $\{\omega_{p}\}$, where $z \in \{1,\ldots,n_z\}$ and $p  \in \{1,\ldots,n_p\}$. All frequencies are normalized to the prototype cutoff frequency $\omega_c$.

\subsubsection{Series inductor transformation}

For transforming a series inductor $g_k s \rightarrow g_k \lambda(s)$, the partial fraction expansion requires computing residues at all poles, including those at $\omega=0$ and $\omega=\infty$ if present. The residues are \cite{guillemin1957synthesis}:
\begin{align}
k_0 &= \begin{cases}
    0 & \text{if } \lambda(0) = 0 \\
    \prod\limits_z \omega_{z}^2 / \prod\limits_p \omega_{p}^2 & \text{if } \lambda(0) = \infty
    \end{cases} \\
k_\infty &= \begin{cases}
    0 & \text{if } \lambda(\infty) = 0 \\
    1 & \text{if } \lambda(\infty) = \infty
    \end{cases}
\end{align}
By convention here, $\prod\limits_z$ is the product over all the zeros, and $\prod\limits_p$ is the product over all the poles. For the residues at finite poles $\omega_{p}$:
\begin{align}
k_p &= \begin{cases}
\dfrac{\prod\limits_z \left(\omega_{z}^2 - \omega_{p}^2\right)}{2 \prod\limits_{m \neq p} \left(\omega_{m}^2 - \omega_{p}^2\right)} \\[2ex]
\qquad \text{if } \lambda(0) = 0\\[3ex]
\dfrac{\prod\limits_z \left(\omega_{z}^2 - \omega_{p}^2\right)}{-2 \omega_{p}^2 \prod\limits_{m \neq p} \left(\omega_{m}^2 - \omega_{p}^2\right)} \\[2ex]
\qquad \text{if } \lambda(0) = \infty
\end{cases}
\end{align}

The transformed circuit elements are, in dimensioned units (Henries and Farads):
\begin{align}
C_0 &= \frac{1}{k_0 g_k Z_0 \omega_c} \\
L_\infty &= \frac{k_\infty g_k Z_0}{\omega_c} \\
L_p &= \frac{2k_p g_k Z_0}{\omega_c \omega_{p}^2}, \quad
C_p = \frac{1}{2k_p g_k Z_0 \omega_c}
\end{align}
where $Z_0$ is the characteristic impedance.

\subsubsection{Shunt capacitor transformation}

For transforming a shunt capacitor $g_k/s \rightarrow g_k/\lambda(s)$, the partial fraction expansion requires computing residues at all zeros, including those at $\omega=0$ and $\omega=\infty$ if present. The residues are \cite{guillemin1957synthesis}:
\begin{align}
k_0 &= \begin{cases}    
    \prod\limits_p \omega_{p}^2 / \prod\limits_z \omega_{z}^2 & \text{if } \lambda(0) = 0 \\
    0 & \text{if } \lambda(0) = \infty
    \end{cases} \\
k_\infty &= \begin{cases}    
    1 & \text{if } \lambda(\infty) = 0 \\
    0 & \text{if } \lambda(\infty) = \infty
    \end{cases}
\end{align}
where $\prod\limits_z$ is the product over all the zeros, and $\prod\limits_p$ is the product over all the poles. For the residues at finite zeros $\omega_{z}$:
\begin{align}
k_z &= \begin{cases}
\dfrac{\prod\limits_p \left(\omega_{p}^2 - \omega_{z}^2\right)}{-2 \omega_{z}^2 \prod\limits_{n \neq z} \left(\omega_{n}^2-\omega_{z}^2\right)} \\[2ex]
\qquad \text{if } \lambda(0) = 0\\[3ex]
\dfrac{\prod\limits_p \left(\omega_{p}^2-\omega_{z}^2\right)}{2 \prod\limits_{n \neq z} \left(\omega_{n}^2-\omega_{z}^2\right)} \\[2ex]
\qquad \text{if } \lambda(0) = \infty
\end{cases}
\end{align}

The transformed circuit elements are, in dimensioned units (Henries and Farads):
\begin{align}
C_0 &= \frac{g_k}{k_0 Z_0 \omega_c} \\
L_\infty &= \frac{k_\infty Z_0}{g_k \omega_c} \\
L_z &= \frac{2k_z Z_0}{g_k \omega_c \omega_{z}^2}, \quad
C_z = \frac{g_k}{2 k_z Z_0 \omega_c}.
\end{align}

\subsection{Foster form 2 (susceptance transformation)}

Consider now a susceptance function $\lambda^{-1}(\omega)$ with non-trivial zeros at $\{\omega_{z}\}$ and poles at $\{\omega_{p}\}$, where $z \in \{1,\ldots,n_z\}$ and $p  \in \{1,\ldots,n_p\}$. Naturally, the zeros and poles of $\lambda^{-1}$ are the poles and zeros of $\lambda$, respectively. Therefore the calculation of the residues for the series inductor case in the Foster form 2 transformation is similar to the calculation of the residues in the shunt capacitor case in the Foster form 1 transformation. Similarly, the calculation of the residues for the shunt capacitor case in the Foster form 2 transformation is similar to the calculation of the residues in the series inductor case in the Foster form 1 transformation. 

Then, for the series inductor case, the transformed circuit elements are, in dimensioned units (Henries and Farads):
\begin{align}
L_0 &= \frac{g_k Z_0}{k_0 \omega_c}\\
C_\infty &= \frac{k_\infty}{g_k Z_0 \omega_c} \\
L_z &= \frac{g_k Z_0}{2 k_z \omega_c}, \quad
C_z = \frac{2 k_z}{g_k Z_0 \omega_c \omega_{z}^2},
\end{align}

while for the shunt capacitor case, the transformed circuit elements are:
\begin{align}
L_0 &= \frac{Z_0}{k_0 g_k \omega_c} \\
C_\infty &= \frac{k_\infty g_k}{Z_0 \omega_c} \\
L_p &= \frac{Z_0}{2 k_p g_k \omega_c}, \quad
C_p = \frac{2 k_p g_k}{Z_0 \omega_c \omega_{p}^2}.
\end{align}

\section{General expression of the wavenumber $k$}
\label{app:kgen}

I consider the ABCD matrix of one supercell of length $d$ within an infinite periodic lattice. Applying Bloch's theorem yields $V(x+d) = e^{-\gamma d} V(x)$, and $I(x+d) = e^{-\gamma d} I(x)$, where $\gamma = \alpha \pm j\beta$, with the sign ambiguity indicating whether one considers forward ($+$) or backward ($-$) propagating waves. In other words,
\begin{equation}
    \begin{pmatrix} A-e^{-\gamma d} & B \\ C & D-e^{-\gamma d} \end{pmatrix} \begin{pmatrix} V(x+d) \\ I(x+d) \end{pmatrix} = 0,
\end{equation}
which has a non-trivial solution if and only if the determinant is null,
\begin{equation}
    \begin{vmatrix} A-e^{-\gamma d} & B \\ C & D-e^{-\gamma d} \end{vmatrix} = 0.
\end{equation}
Using the reciprocity condition $AD-BC=1$ this determinant yields
\begin{equation}
    \cosh(\gamma d) = \frac{A+D}{2}.
\end{equation}
When $\alpha=0$ (lossless ATL), it yields
\begin{equation}
    \beta =  \pm\frac{1}{d}\Im{\arccosh\left(\frac{A+D}{2}\right)},
\end{equation}
which, given the sign selection rule for $v_g>0$ described in Sec.\,\ref{sec:LPATL} yields Eq.\,\ref{eq:k} (for $d=1$).

\section{Foster's reactance theorem in $k$-space}
\label{app:kFoster}

I establish that in any lossless ATL, the wavenumber $k$ is a strictly monotonically increasing function of frequency $\omega$ within each passband. This property, analogous to Foster's reactance theorem, guarantees the absence of anomalous dispersion and ensures a unique correspondence between frequency and wavenumber for forward-propagating modes.

I first prove this for a low-pass prototype (\ie{}, LC ladder). The wavenumber $k$ represents the negative phase shift per unit cell, constrained to the irreducible Brillouin zone (IBZ) $[0,\pi]$ due to lattice periodicity \cite{hwang2012periodic}. From Eq.\,\ref{eq:k}, $k$ is determined by the half-trace of the unit cell ABCD matrix:
\begin{equation}
    T = \frac{A+D}{2},
\end{equation}
which is a polynomial in $\omega$ with real coefficients. Within the IBZ, when $|T| \leq 1$ it yields:
\begin{equation}
     k = \arccos(T),
\label{eq:kpass}
\end{equation}
indicating a passband. When $|T| > 1$, $\arccosh(T)$ has a positive real part, corresponding to an evanescent mode (stopband).

\begin{figure}[h!]
\includegraphics[width=\columnwidth]{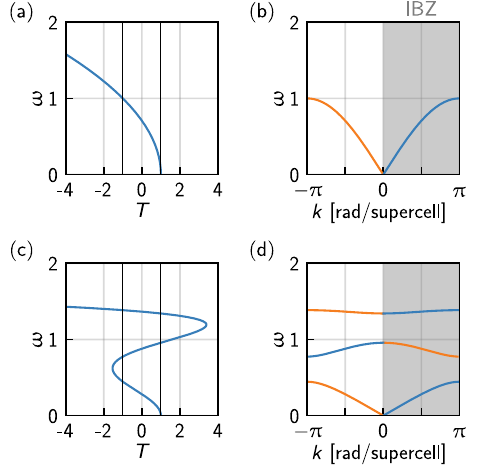}
\caption{Half-trace $T$ and corresponding band diagrams for (a) an unloaded ATL and (b) a periodically loaded ATL. The extrema of $T$ (where $T' = 0$) occur in stopband regions where $|T| \geq 1$, ensuring $T$ remains monotonic within each passband.}
\label{fig:phase}
\end{figure}

The behavior of $T$ determines $k$ within the IBZ, as illustrated in Fig.\,\ref{fig:phase}. For a lossless, reciprocal network, the ABCD eigenvalues are
\begin{equation}
    m_\pm = T\pm j\sqrt{1-T^2},
\end{equation}
and \textit{in the passband} they lie on the unit circle: $\lvert m_\pm \rvert = 1$. This implies the discriminant of the square root to be positive, which gives
\begin{equation}
    m_\pm = \cos k \pm j\sin k = e^{\pm jk},
\end{equation}
with $\sin k \geq 0$. Differentiating $T=\cos k$, the group velocity becomes:
\begin{equation}
    v_g = \frac{d\omega}{dk} = \frac{-\sin{k}}{T'},
\end{equation}
where $T' = dT/d\omega$.

The group velocity represents the speed at which a signal propagates within the ATL, therefore it must not diverge within passbands, so $T' \neq 0$ when $|T| < 1$. In other words, for a lossless LC ladder, the extrema of $T$ (where $T' = 0$) occur when $|T| \geq 1$, \ie{}, in stopband regions. This ensures $T$ is strictly monotonic within each passband. Since $k = \arccos(T)$ and arccos is a decreasing function, if $T$ decreases monotonically ($T' < 0$), then $k$ increases monotonically ($v_g > 0$). Conversely, bands with $T' > 0$ correspond to $v_g < 0$. Taking the forward-propagating branch ($v_g > 0$) ensures $k$ is always a monotonically increasing function of $\omega$.

This theorem extends to frequency-transformed ATLs. Since any realizable reactance $\lambda(\omega)$ or susceptance $\lambda^{-1}(\omega)$ is monotonically increasing \cite{foster1924a}, the composition $k(\lambda(\omega))$ or $k(\lambda^{-1}(\omega))$ preserves monotonicity. This guarantees well-behaved dispersion even in complex filter-based ATL designs.

\section{Inductance-flux Taylor expansion of a kinetic-inductance material}
\label{app:Lphi}

I derive the expression of the kinetic inductance $L$ as a function of the branch flux $\varphi$. Starting from $L$ expressed as a Taylor expansion in the
  total current $I$ \cite{Zmuidzinas2012superconducting}
\begin{equation}
    L(I) = L_0\left[1 + \left(\frac{I}{I_*}\right)^2\right],
\end{equation}
when $I$ contains a dc component $I_d$, $L(I)$ becomes \cite{malnou2021three}
\begin{equation}
    L(I) = L_d\left(1+\epsilon I + \xi I^2\right),
\label{eq:LI}
\end{equation}
where $L_d=L_0(1+I_d^2/I_*^2)$, $\epsilon = 2I_d/(I_*^2+I_d^2)$, and $\xi = 1/(I_*^2+I_d^2)$.

The kinetic inductance can alternatively be expressed through the constitutive relation:
\begin{equation}
    V = \varphi_0\frac{d\varphi}{dt} = L \frac{dI}{dt},
\label{eq:VdIdt}
\end{equation}
where $\varphi_0$ is the reduced flux quantum and $\varphi$ is the normalized branch flux. From this relation, $L$ becomes:
\begin{equation}
    L = \varphi_0\frac{d\varphi}{dI}.
\label{eq:LdphidI}
\end{equation}
Using $L(I)$ defined in Eq.\,\ref{eq:LI}, integrating both sides of Eq.\,\ref{eq:VdIdt} yields
\begin{equation}
    \varphi = \frac{L_d}{\varphi_0}\left(I + \frac{\epsilon}{2}I^2 + \frac{\xi}{3}I^3\right).
\end{equation}
To obtain the current as a function of branch flux, this series is inverted:
\begin{equation}
    I = \frac{\varphi_0}{L_d}\left[\varphi - \frac{\varphi_0}{L_d}\frac{\epsilon}{2}\varphi^2 + \left(\frac{\varphi_0}{L_d}\right)^2\left(\frac{\epsilon^2}{2}-\frac{\xi}{3}\right)\varphi^3\right].
\end{equation}
Differentiating $I(\varphi)$ and substituting $dI/d\varphi$ into Eq.\,\ref{eq:LdphidI} yields $L(\varphi)$:
\begin{equation}
    L = \frac{L_d}{1-\frac{\varphi_0}{L_d}\epsilon\varphi + \left(\frac{\varphi_0}{L_d}\right)^2\left(\frac{3\epsilon^2}{2}-\xi\right)\varphi^2},
\end{equation}
giving the expansion:
\begin{equation}
    L = L_d\left[1+I_0\epsilon\varphi - I_0^2\left(\frac{3\epsilon^2}{2}-\xi\right)\varphi^2\right],
\end{equation}
where $I_0=\varphi_0/L_d$ is the inductor's characteristic current. For $I_d=0$, this reduces to:
\begin{equation}
    L = L_0\left[1+\left(\frac{I_0}{I_*}\varphi\right)^2\right],
\end{equation}
where $I_0=\varphi_0/L_0$.

\section{Component values for TWPA designs}
\label{app:values}

\begin{figure}[h!]
\includegraphics[width=\columnwidth]{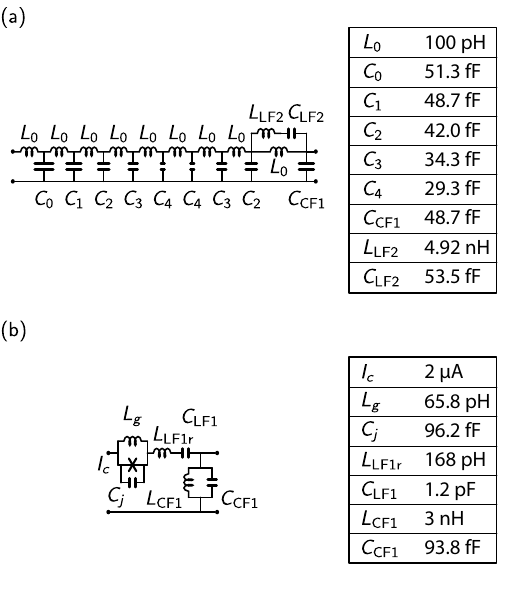}
\caption{Circuit component values within the supercell of the 4WM KTWPA (a), and within the b-TWPA cell (b).}
\label{fig:twpavalues}
\end{figure}

Figure \ref{fig:twpavalues} provides the component values for both TWPA designs presented in Sec.\,\ref{sec:examples}. These values result from applying the synthesis procedures developed in this work. Panel (a) shows the 4WM KTWPA supercell, featuring periodic capacitor modulation ($C_0$ through $C_4$) combined with a resonant phase-matching (rpm) filter ($L_\mathrm{LF2}$, $C_\mathrm{LF2}$). The base inductance $L_0 = 100$\,pH and the periodic modulation create the desired stopband structure, while the rpm filter provides sharp phase accumulation for 4WM phase matching. Panel (b) presents the b-TWPA unit cell with its ambidextrous design. The series LC resonator ($L_\mathrm{LF1r}$, $C_\mathrm{LF1}$) and parallel LC resonator ($L_\mathrm{CF1}$, $C_\mathrm{CF1}$) create the composite right-left-handed behavior, while the biased rf-SQUID (characterized by critical current $I_c = 2$\,\micro A and geometric inductance $L_g = 65.8$\,pH) provides the nonlinearity for 3WM parametric amplification. All designs target a $50$\,\ohm{} characteristic impedance.

\section{\cor{Third-harmonic suppression in the 4WM KTWPA}}
\label{app:thirdharmonic}

\begin{figure}[h!]
\includegraphics[width=\columnwidth]{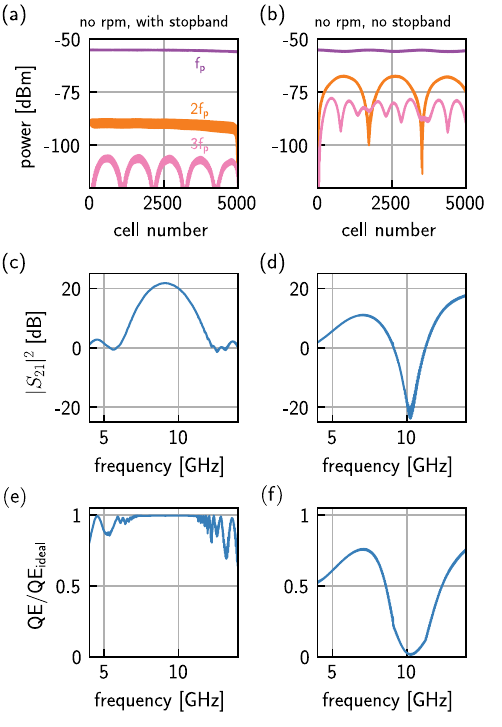}
\caption{\cor{Effect of the third-harmonic stopband on the KTWPA of Sec.\,\ref{sec:4WMKTWPA}, with the rpm features removed. The left column (a,c,e) corresponds to the design with the stopband at $3f_p$, and the right column (b,d,f) to the design without. (a,b) Spatial power distribution of the pump fundamental ($f_p$, purple), second harmonic ($2f_p$, orange), and third harmonic ($3f_p$, pink) along the device. Without the stopband, the 2nd and 3rd harmonics reach peak powers 20–27\,dB higher. (c,d) Signal gain $|S_{21}|^2$. With the stopband, the gain profile is symmetric about the pump frequency, characteristic of the 4WM process. Without it, the profile becomes strongly asymmetric, indicating that parasitic higher-order parametric processes dominate the amplification. (e,f) Quantum efficiency normalized to the theoretical maximum. With the stopband, $\mathrm{QE}/\mathrm{QE_{ideal}}$ remains above $0.95$ across the signal band; without it, it degrades to $0.52$–$0.76$.}}
\label{fig:thirdharmonic}
\end{figure}

\cor{Figure \ref{fig:thirdharmonic} compares two variants of the KTWPA presented in Sec.\,\ref{sec:4WMKTWPA}, with the rpm features removed to isolate the effect of the third-harmonic stopband: one retaining the stopband at $3f_p$ and one without. Without the stopband, the 2nd and 3rd pump harmonics reach peak powers of $-67$\,dBm and $-78$\,dBm respectively, compared to $-88$\,dBm and $-105$\,dBm with the stopband, a suppression of approximately 20–27\,dB. Both designs exhibit similar pump power variations ($\sim28\,\%$) along the line, indicating that the stopband does not primarily prevent pump depletion. Rather, its essential role is to enforce that amplification occurs exclusively through the designed 4WM process: without the stopband, the signal gain profile becomes strongly asymmetric about the pump frequency, and the quantum efficiency degrades from $>0.95$ to $0.52$–$0.76$.}

\clearpage


%

\end{document}